\documentclass[12pt, oneside]{article}   	
\usepackage{graphicx}						
\usepackage{amssymb}
\usepackage{amsmath}
\usepackage{titlesec}
\usepackage{algorithm}
\usepackage{algpseudocode}

\title{Forecasting Extreme Temperatures in Siberia Using Supervised Learning and Conformal Prediction Regions.}
\author{Richard Berk, University of Pennsylvania \\
Amy Braverman, The Jet Propulsion Laboratory \\ and California Institute of Technology}
\date{\today}

\begin{document}
\maketitle

\begin{abstract}
In this paper, we step back from a variety of competing heat wave definitions and forecast directly unusually high temperatures. Our testbed is the Russian Far East in the summers of 2022 and 2023. Remotely sensed data from NASA's Aqua spacecraft are organized into a within-subject design that can reduce nuisance variation in forecasted temperatures. Spatial grid cells are the study units. Each is exposed to precursors of a faux heat wave in 2022 and to precursors of a reported heat wave in 2023. The precursors are used to forecast temperatures two weeks in the future for each of 31 consecutive days. Algorithmic fitting procedures produce forecasts with promise and relatively small conformal prediction regions having a coverage probability of at least .75. Spatial and temporal dependence are manageable. At worst, there is weak dependence such that conformal prediction inference is only asymptotically valid.
\end{abstract}


\section{Introduction}

For more than a decade, the number research publications on extreme weather events has been growing. The study of heat waves and their consequences has been a major focus (Perkins, 2015; Pitcar et al., 2019; Marx et al., 2020; Klingh\"ofer et al., 2023). With a very few exceptions (Russo et al., 2014; Jaque-Dumas, 2022; Khantana et al., 2024), heat wave forecasting has not been the organizing motivation. Description and explanation have dominated the narrative (Mann, 2018; Mckinnon and Simpson, 2022; Petoukhov et al., 2022; Li et al, 2024). Moreover, much of the forecasting work has been constrained by reliance on simulation output rather than primary data, challenges inherent in the study of rare events, insufficiently developed uncertainty quantification, and conceptual ambiguities surrounding heat wave definitions. 

In this paper, we take a step back. We forecast extreme temperatures without being conceptually limited to a predetermined heat wave definition. In this manner, we seek to sidestep the lack of consensus about how heat waves should be operationalized. Among the benefits, we have no need to impose off-the-shelf statistical distributions, such as a re-centered normal or the generalized extreme value formulation. We can favor nonparametric approaches unless the data counsel otherwise. We also have no need to commit to heat waves as discrete events, which is at least implicitly very common. In short, our approach is substantially data-driven. 

We draw on the work of Berk and his colleagues (Berk et al., 2024), who use remotely sensed AIRS data to lay a foundation for forecasts linked to reported heat waves in the Pacific Northwest and the greater Phoenix area. We also rely heavily on machine learning procedures that can work well with conformal prediction regions; we offer valid prediction intervals with no distributional assumptions. Finally, we use as a testbed the Russian far east. Its enormous size and topographical complexity present forecasting challenges, and the warming underway may turn out to be catastrophic (Witze, 2020; Callaghan et al., 2021; Kim et al., 2022).

In section 2, AIRS data for the Russian far east are described, including the use of remotely sensed grid cells as the study units. Section 3, addresses how the data were  curated and explored to produce a within-subject research design that can reduce the amount of ``noise'' in proper forecasts. Construction of the response variable is also a major theme. Section 4 contains the data analysis, which relies substantially on random forests, quantile random forests, and loess smoothers. Potential complications from temporal and spatial dependence are considered.  Forecasting results are presented along with conformal prediction regions that provide valid uncertainty assessments.  Section 5 elaborates on the findings, discusses the endgenous sampling problem, expands on the challenges from missing data, and considers how forecasts of extreme high temperatures need to be conceptually connected to heat wave definitions. In section 6, the overall results are succinctly summarized and next steps briefly anticipated. 
\section{Data}

Our central goal is to forecast two weeks in advance unusually high temperatures, often treated as the signature feature of extreme heat waves. We use the Russian far east as our study site. The early June, 2023 reported heat wave, centered in Yakutia and Irkutsk regions, serves as the empirical focus. We stress the word ``reported'', taking no position on the accuracy of all claims made. Conditional on time and place, temperature variability tends to be noticed when perceived as unusual, especially if it is worrisome as well. Temperatures can become, therefore, newsworthy as well as noteworthy. Although grounded in observable physical phenomena, reported heat waves can be usefully considered social constructions (Hulme et al., 2008; Hopke, 2019) whose available details should be treated with some scientific suspicion. We stress that of all the locations and dates around the globe where temperatures variation could be studied, we use the Siberian 2023 reported heat wave solely as a means to locate when and where instructive data might be found. 

The approximate boundaries of the reported heat wave are latitude $50^\circ$ N  to $65^\circ$ N and longitude $75^\circ$ E to $140^\circ$ E. The data we use were collected and initially processed by NASA’s Atmospheric Infrared Sounder (AIRS) instrument aboard the Aqua spacecraft (Aumann et al., 2003), described in more detail elsewhere (Tien, 2020). The Aqua spacecraft was launched into a polar orbit at an altitude of 705 km on May 4th, 2002. Each grid cell is measured toward the middle of each day once every 24 hours.

We begin with grid cells as the study units. They are rectangles that are very close to squares closer to the equator. Farther north, they become increasingly elongated in the north-south direction. Each grid cell encompasses about 1000 square km.  Radiation from each entire grid cell is sensed and processed into measures of interest to climate scientists, bundled for the grid cell as a whole. Grid cell location is defined by the longitude and latitude of the grid cell centroid. Distances between grid cells can be computed from the centroid locations, which makes empirical assessments of spatial dependence challenging. For example, grid cells that share a common border can, nevertheless, have empirical distances between centroids of at least about 100 km.  

An important characteristic of some variables is that they are provided to researchers at different distances above sea level, nominally levels 1 through 12. Smaller numbers mean that the measure is taken closer to sea level.  However, because grid cells can be at different elevations, readings near sea level often cannot be obtained. Only radiation from ground level and above can be received by the satellite.\footnote
{
The ``level'' of a variable is determined by atmospheric pressure that is higher at lower altitudes.  The units of pressure are hpa (i.e., hectopascale), one of which is equivalent to 100 pascals (Pa). A pascal is a unit of force based on Newton's second low. One newton is defined as the amount of force needed to accelerate a one-kilogram mass at a rate of one meter per second squared. Level 8, used often in the analyses to follow, is located at an altitude of about 9,000 m (Susskind et al., 2014, Tian et al., 2020). 
}

\section{Research Design}

\begin{figure}[htbp]
\begin{center}
\includegraphics[width=5.5in]{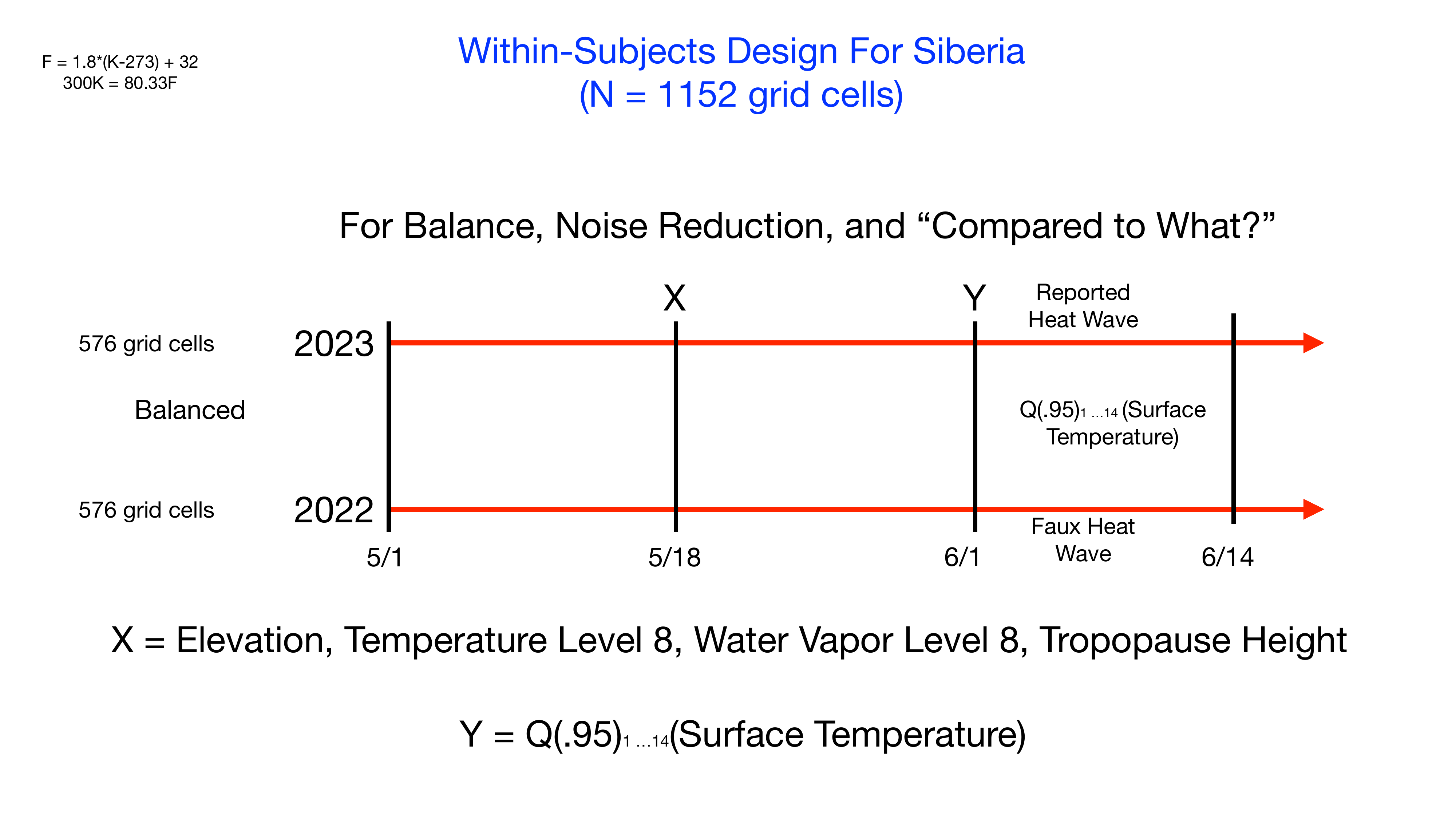}
\caption{Within-Subject Design: Each grid cell is exposed to reported heat wave precursors and to faux heat wave precursors. Y is the response variable computed as the Q(.95) value of each grid cell over the first two weeks of the reported and faux heat wave for surface temperatures in Kelvin units. X includes the four precursors from approximately two weeks earlier. For readers who think in Fahrenheit, the conversion from Kelvin units is shown on the upper left corner of the figure with an example.}
\label{fig:design}
\end{center}
\end{figure}

Figure~\ref{fig:design} shows our within-subject design (Maxwell et al., 2018: Section III). The layout in 2023 is duplicated in 2022. The same grid cells were exposed on different dates to the reported heat wave precursors and to the faux heat wave precursors. The faux heat wave was constructed from 2022 Siberian data, absent a reported heat wave, having the exact same structure as the data for the reported heat wave one year earlier. It is meant to be one possible answer to the question ``extreme temperatures compared to what?" The faux heat data can be seen as the product of a comparison intervention (Maxwell et al.,, 2018: Section III), and will be very useful later introducing additional, systematic temperature variability. Further discussion of other possible comparisons is provided subsequently. 

Another asset of the design is that if one set of grid cells were exposed to the reported heat wave precursors and another set of grid cells were exposed to the faux heat wave precursors, fixed features of the grid cells related to the two different spatial locations would introduce additional nuisance variability into the forecasts; ``.. because comparisons in the repeat-measures design are made within subjects, variability in individual differences between subjects is removed from the error term'' (Maxwell et al., 2018: 614). 
 
Finally, it is well known that reported heat waves are rare events. If for any day anywhere on the planet, one forecasted no unusually warm temperatures, that forecast would almost certainly be correct. The same applies the Russian Far East, even if only the summer months are considered. In other words, there would be very few, if any, unusually warm, temperatures in the dataset making the study of extreme temperatures daunting. Our design responds by approximately balancing the data such that the number of grid cells exposed precursors of a reported heat is about the same as the number of grid cells exposed to the precursors of a faux heat wave. However, this strategy introduces endogenous sampling that over-represents likelihood of unusually high temperatures compared to their presence in most real-world populations. As a result, statistical models or algorithms risk producing serious distortions (Manski and Lerman, 1977; Manski and McFadden, 1981). The issues can be subtle and will be addressed in some depth after the forecasting results are discussed.

Precursor variables for both the reported heat wave and the faux heat wave include (1) grid cell elevation, (2) temperature at level 8 in Kelvin, (3) water vapor concentrations at level 8, and (4) the height of the tropopause. These variables, measured in the summer 2022 again in the summer of 2023, were productively employed in a demonstration-of-concept paper by Berk and his colleagues (2024). We also duplicate their balance between reported heat wave observations and faux heat wave observations.\footnote
{
Water vapor concentration is a lot like relative humidity at different altitudes. It is the proportion of the total atmospheric profile that is water vapor at a given altitude. The total atmospheric profile is the proportional presence of different constituents of the atmosphere at a given altitude such as oxygen, carbon dioxide, nitrogen, and water vapor. The tropopause is the boundary layer between the troposphere and the stratosphere. It is the point at which the temperature stops decreasing with altitude and begins to increase. The height of the tropopause varies with season, spatial location and local weather conditions. The last can be relevant to unusually high temperatures.
}

Ideally, our response variable should be one or more different functions of ground surface temperature in Kelvin units. However, we are disadvantaged by substantial missing data on all variables below level 4. Grid cells near sea level may be the explanation. If those grid cells are at higher elevations, no radiation closer to sea level is available to the satellite; available radiation begins at ground level. We use surface temperature nevertheless because similar missing data problems are found for precursor tropopause height. Surface temperature has the interpretative advantage because much of ecosystem damage occurs at ground level. We return to missing data issues later.\footnote 
{
We considered using for the response variable grid cell surface temperatures supplemented with imputed values for missing surface temperatures using temperature at level 4 (which was reasonably complete). But imputing values for a response variable from the anticipated predictors, or variables related to the anticipated predictors, is ill-advised. Associations between the predictors and the response are being artificially build in. Moreover, recent work indicates that even powerful nonlinear imputation procedure often do not improve fitting performance enough to matter (Le Morvan and Varoquaux, 2025).
}

We settled on the 0.95 quantile as an easily understood measure of ``unusual.'' Alternative large quantiles might have served as well. Very different measures might be preferred going forward as the science of extreme heat effects on ecosystems advances. 

A significant feature of the design is the temporal lag imposed on the precursors. As a matter of scientific interest and subsequent public policy, forecasts well in advance are important. For example, a two-week gap can give local first-responders such as fire fighters  time to prepare. But, two weeks before what? Given our interest in unusually high temperatures, we settled on trying to find a lag that targeted the Q(.95) surface temperature over of each grid cell during each of the first 14 days of the reported heat wave; there needed to be one Q(.95) value for each day. Weather forecasts two weeks in advance produced by machine learning are becoming increasingly accurate and practical (Price et al., 2025).

\begin{figure}[htbp]
\begin{center}
\includegraphics[width=2.5in]{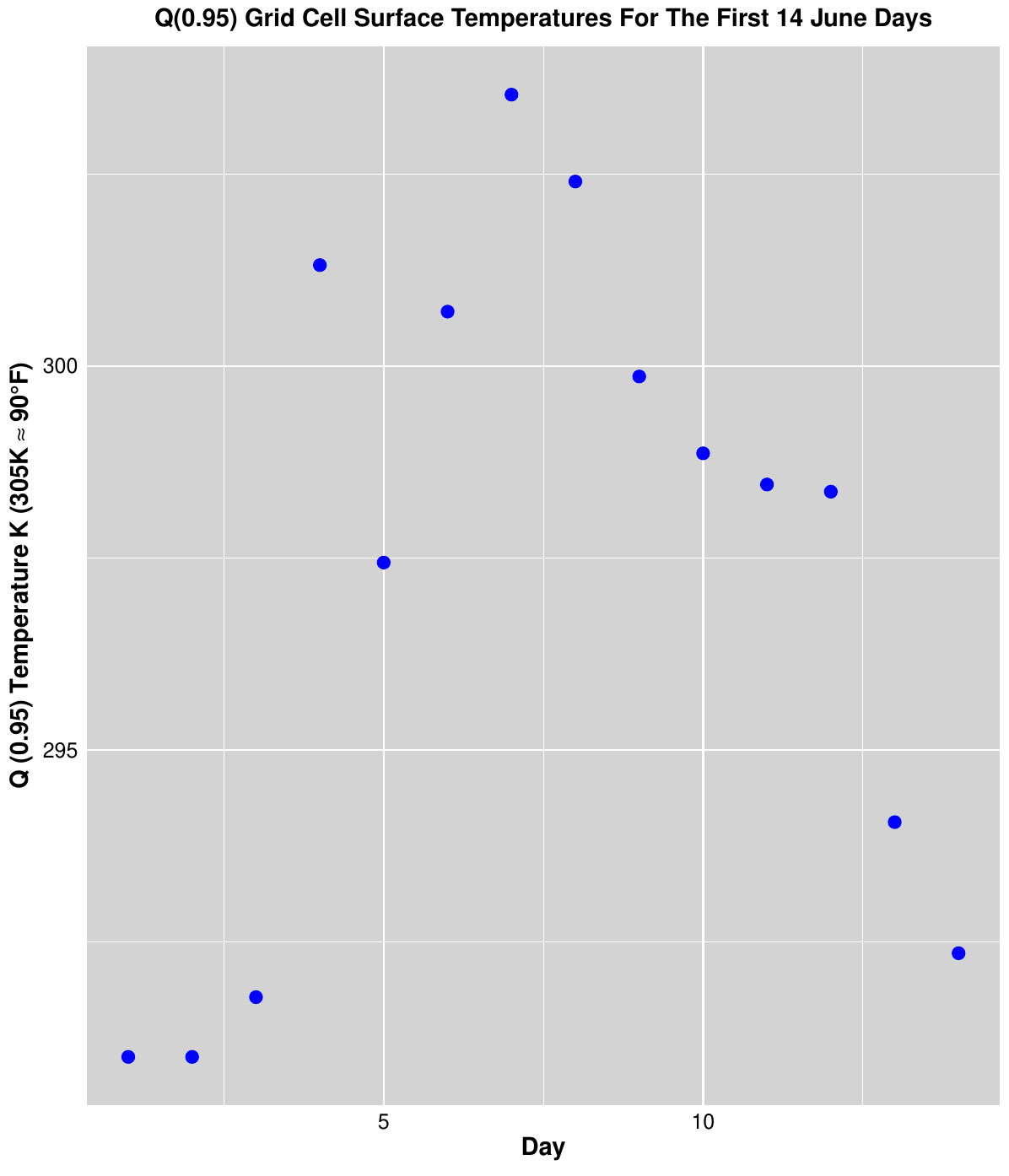}
\caption{Within-Subject Design: Q(0.95) observed surface temperature values day by day over grid cells for each of the first 14 days in June.}
\label{fig:lag}
\end{center}
\end{figure}

Figure~\ref{fig:lag} shows the Q(.95) surface temperatures over all grid cells for each of the first 14 days of the reported Siberian heat wave beginning on or about June 1, 2023. There is an increasing trend for the first week followed by a gradual decline. The highest Q(.95) temperature across grid cells day by day occurs on June 7th with other high temperatures occurring from days 5 through 8. If such a pattern were known when forecasts were needed, building in a 14 day lead going forward from the precursors might require using May 19th for the day on which the precursors were measured for both the reported heat wave, and for design comparability, the faux heat wave.  But in practice, such information would not be available. It seems prudent to set the precursor measurement day one day earlier to the 18th because on that date two weeks later the warming increases appeared to begin. 

However, a single grid cell's Q(.95) surface temperature might fall anywhere in the first two heat wave weeks (or later). In other words, the actual lead time may turn out empirically to be up to a month (or more) for particular grid cells. We will see shortly how this plays out. But clearly, the lead time provided is a rough approximation at the level of days.

Anticipating forecasts to come, it may be worth noting now that Figure~\ref{fig:lag} offers a rough facsimile of an estimated concave parabola for the first two weeks of the reported Siberian heat wave. With our focus so far on data to be used in algorithm training, one might expect something of this form to serve as a sanity check when day to day forecasts are soon provided.

To summarize, the data were stacked such that half of the observations are associated with reported heat wave and half of the observations are associated with the faux heat wave. The within-subject design is maintained. After deletions for missing data, the stacked dataset contains 877 observations. The major missing-data predictor was tropopause height. Concerns about the representativeness of the complete-case stacked data are addressed later after the forecasting results are presented. 

\section{Data Analysis}

Random forests was applied to the 887 stacked observation with the response variable as each grid cell's Q(.95) surface temperature over the first two weeks in June.\footnote
{
Intermittently, we tried two other supervised learning procedures: stochastic gradient boosting and deep neural networks. Stochastic gradient boosting performed as well as random forests, but deep neural networks performed a bit worse. It is difficult say anything conclusive, however, because small differences, and even some with substantial differences, can be attributed tuning decisions.
}
Predictors were the precursors 14 days earlier. We wrote the code so that the data could be weighted to adjust for endogenous sampling. But for now, all observations are given the same weight. The standard defaults worked well. 

\subsection{The Random Forests Fit for Grid Cells}

\begin{figure}[htbp]
\begin{center}
\includegraphics[width=2.5in]{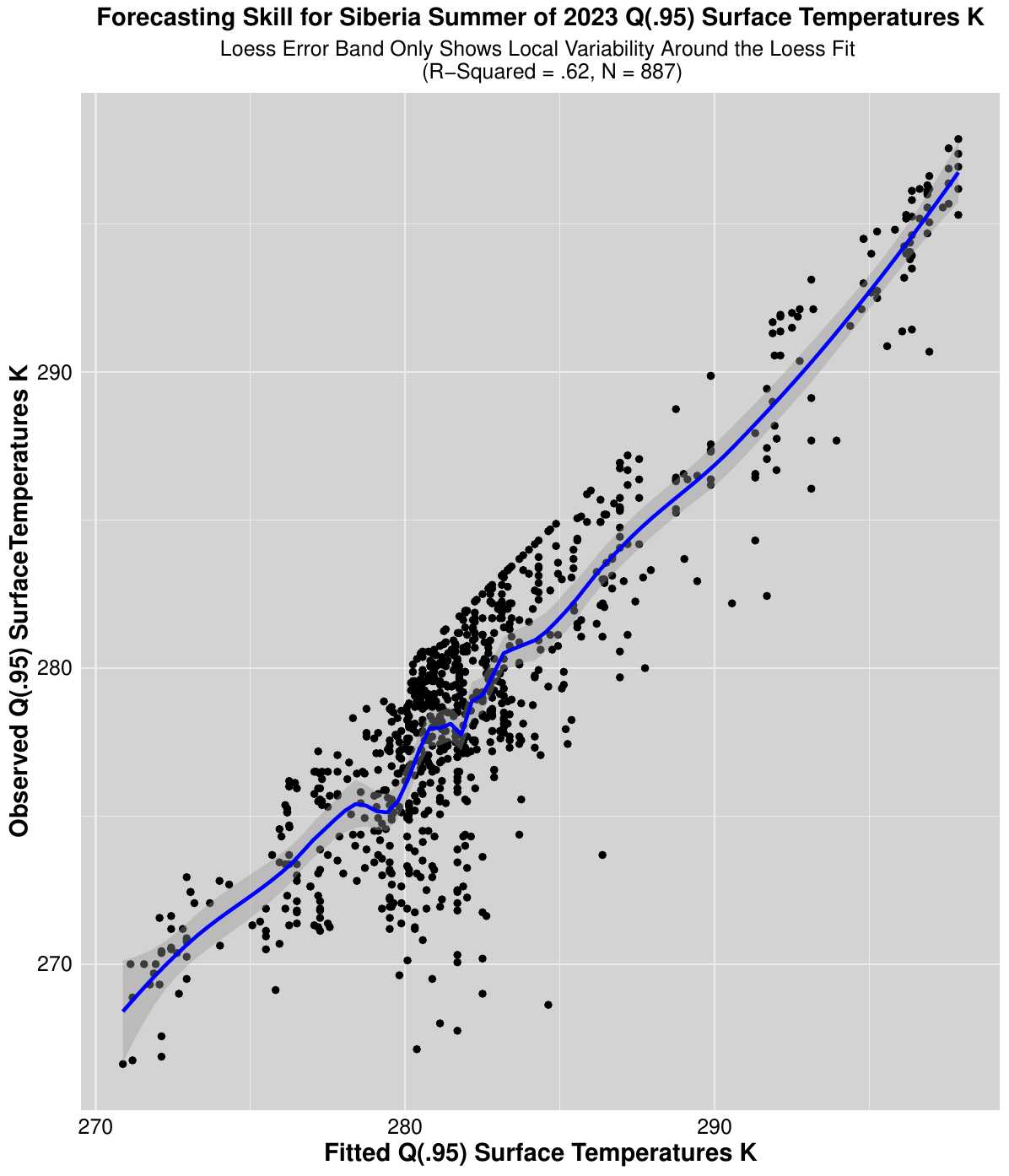}
\caption{Observed Q(.95) response variable values plotted against random forest fitted response variable values with a Loess smoother overlaid.}
\label{fig:fitquality}
\end{center}
\end{figure}

Using the four precursors described earlier, random forests was able to account for 62\% of the variance in the observed Q(.95) grid cell surface temperatures despite the data quality problems. Figure~\ref{fig:fitquality} shows a comforting relationship between the observed Q(.95) surface temperatures plotted against the random forest fitted values.  With all  887 grid cells represented, the loess (nonparametric) smoother in blue displays a near linear relationship (Jacoby, 2000).  The gray bands are the conventional 2 standard error fitting intervals that reinforce a perception of near linearity. But they do not posses valid coverage probabilities because of major assumption violations, and they are \emph{not} forecasting regions. The Y-values are known and included in the data. There is no need to forecast them. The smoothed values in blue have an uncomplicated structure. Because the observed Q(.95)  temperatures and the fitted Q(.95) temperatures are supposed to be quantifying the same thing, the simple relationship is encouraging. 

\begin{figure}[htbp]
\begin{center}
\includegraphics[width=2.5in]{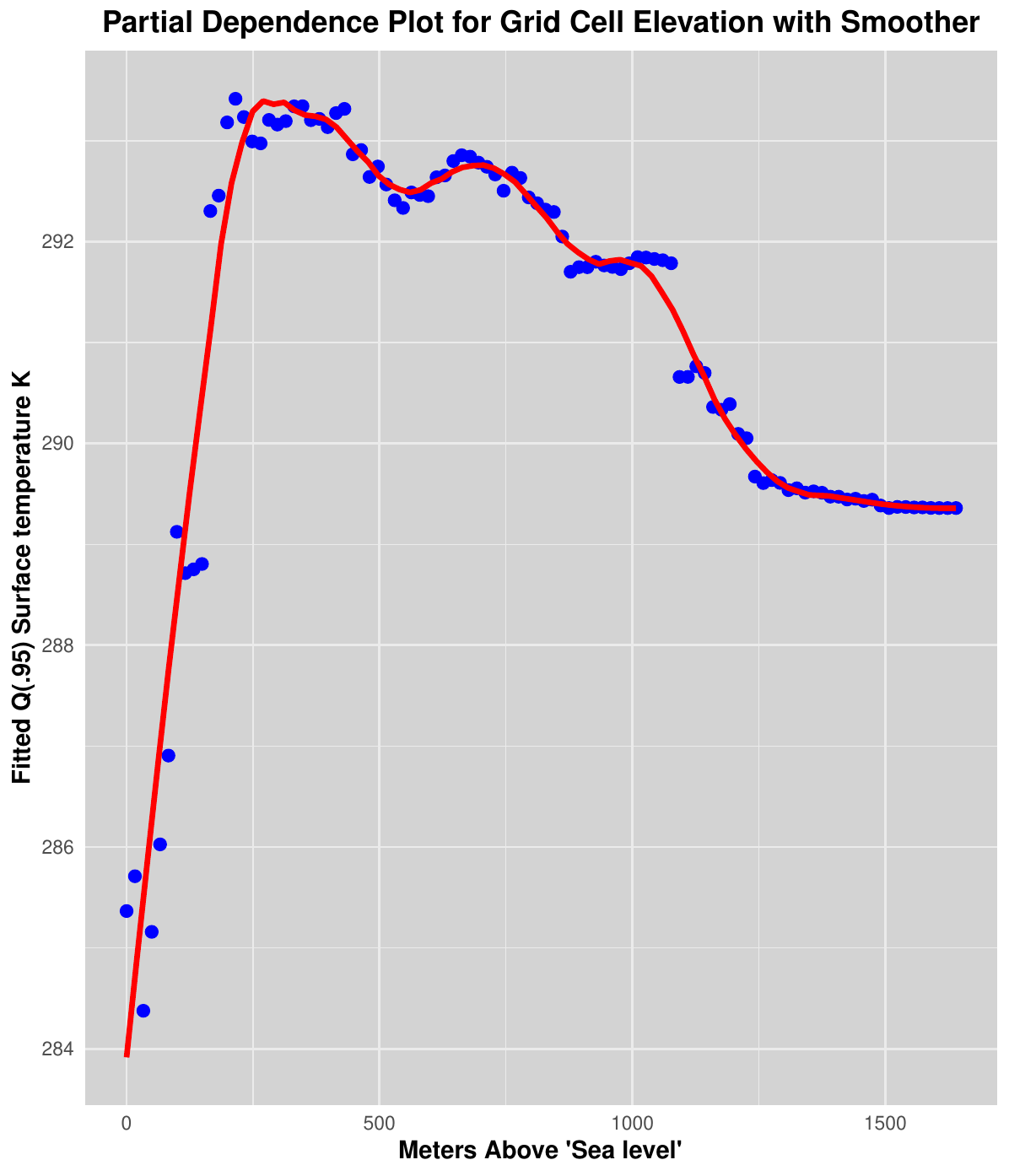}
\caption{Partial dependence plot for grid cell elevation. For this plot, The other three precursors are fixed at their mean values.}
\label{fig:pdp}
\end{center}
\end{figure}

Random forests is not a model, and if it were, the mean function would be badly misspecified because of omitted variables alone. Still, it may be of some interest that  Figure~\ref{fig:pdp} shows that precursor grid cell elevation has a partial dependence plot that is highly nonlinear. The partial dependence plots for the other three predictors are also nonlinear. Perhaps the major point not apparent from Figure~\ref{fig:fitquality} is that at the level of grid cells, Q(.95) surface temperature is related in a nonlinear fashion to each of the four predictors measured at least 2 weeks earlier.\footnote
{
For a numeric response variable, a partial dependence plot displays the relationship between the fitted mean of that response variable, conditional on the values of a particular predictor, with all other predictors fixed at their means. To improve computational speed, the values of the single predictor are usually binned. 
} 

\subsection{Grid Cell Spatial Dependence}

One has in hand from the random forests' fitted values that are the building blocks for forecasting any of the 887 grid cell Q(.95) surface temperatures if subsequently, new cases are realized from the same climate processes, absent their observed Y-values. We consider shortly whether as a policy or scientific matter one should care about Q(.95) temperature forecasts for individual grid cells. 

In principle, conformal prediction regions can be constructed for individual grid cells forecasts. Should forecasting uncertainty be a concern at that granular level, a major requirement is that random forest \emph{residuals} are exchangeable. The most apparent obstacle for these data is residuals that evidence spatial dependence. 

Most of the standard statistical tools for examining spatial dependence are for these data challenging to apply because the within-subject design mandates ideally that all grid cells are included in the data \emph{twice}. As a result, the distance between them is 0.0 km. This can cause problems for many kinds of spatial calculations. More fundamentally, spatial dependence methods generally are built on Tobler's ``First Law of Geography,'' which assumes that closer observations in space are more alike and have higher spatial correlations (Harvey, 2004). Yet, because the precursor values differ in 2023 compared to 2022, and because Q(.95) temperatures are random variables, some of the paired residuals, necessarily at the exact same location, could be rather different. 

Moreover, recall that even some grid cells that share the same border will be estimated to be at least 100 km apart. One consequence is that problematic spatial dependence must have a very long reach: often well over 1000 km between grid cell centroids. This may be an unreasonable expectation although extreme heat caused by heat domes can cover very large areas.

\begin{figure}[htbp]
\begin{center}
\includegraphics[width=2.5in]{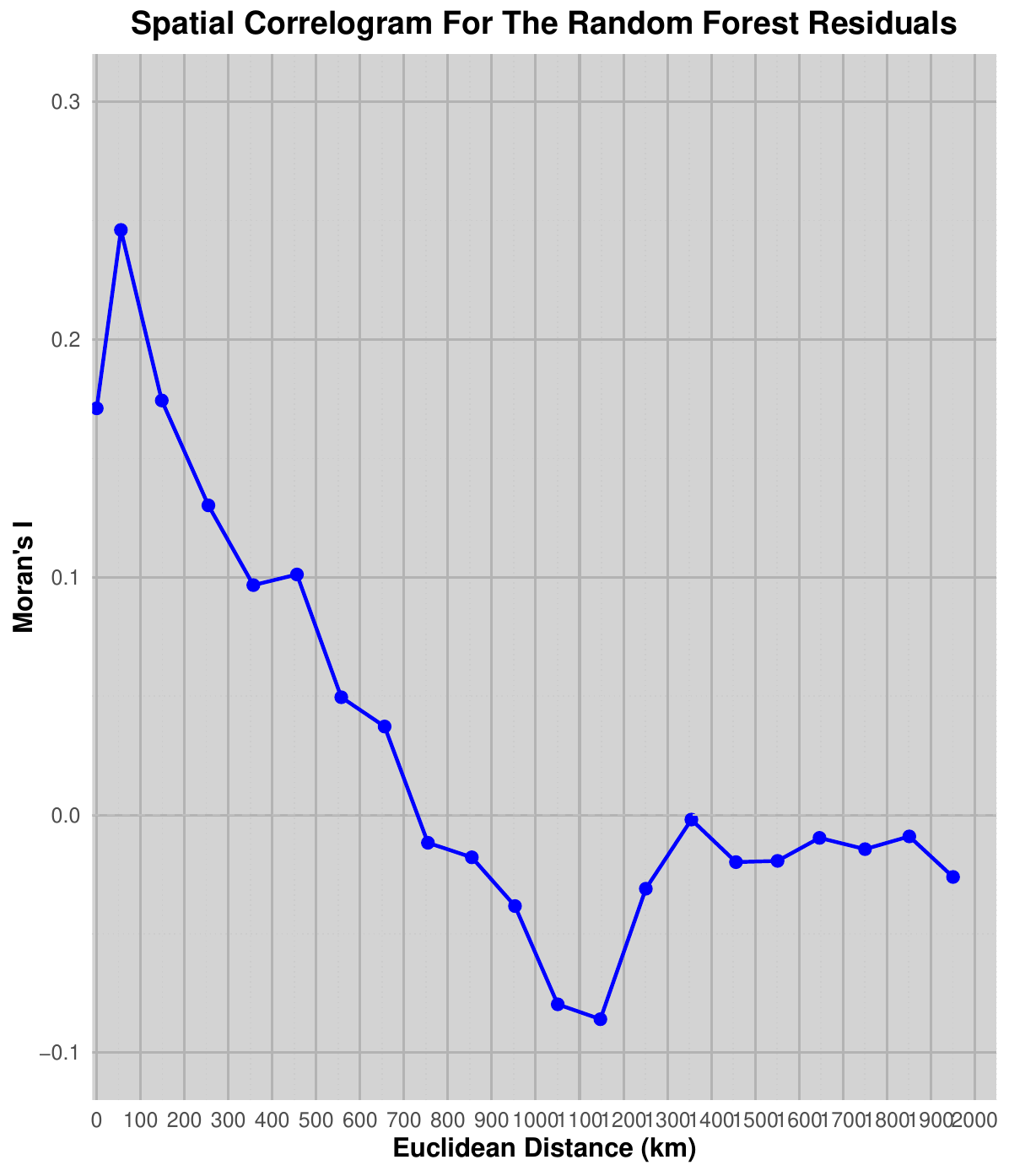}
\caption{Correlogram for the residuals from the random forest fit (N = 887).}
\label{fig:correlogram}
\end{center}
\end{figure}

Figure~\ref{fig:correlogram} shows a correlogram using Moran's I. It is at least curious that the largest value for Moran's I (i.e., about .25) occurs at distance of 150 km between grid cell centroids, while the correlation between centroids at the \emph{same} location is about .09. There is a clear violation of Tobler's First Law of Geography. The violation could be real or an artifact produced by the within-subjects design. 

If, nevertheless, one wants to take Figure~\ref{fig:correlogram} seriously, the spatial correlations are quite small and decline sharply beyond distances of 150 km. Soon the correlations vary between about  .10 and -.10. Within that range, one cannot properly reject at the nominal .05 level the usual null hypothesis that Moran's I equals 0.0. If smaller correlations increasingly appear at greater distances, ideally tapering off to approximately 0.0, a case can be made for weak dependence in the residuals. Then, conformal prediction regions are be justified asymptotically even though finite sample guarantees are lost (Mao et al., 2022; Jiang and Xie, 20224, Oliveira et al., 2024). 

Assuming that one might be interested in forecasting using conformal prediction regions for any of the 887 grid cells, we constructed conformal prediction intervals for several illustrative cases. After fitting, we subsetted the data to include the highest 25\% of the fitted values because it would the higher surface temperature forecasts in which there would be most interest. Moreover, the variance around the fit was not constant. The larger scatter toward the middle of fitted values was a nuisance to avoid. Stipulating weak dependence and an a priori coverage probability of .75, the prediction intervals were generally bounded  by $\pm$ about one Kevin unit.\footnote
{
If all of the data are included, the prediction intervals are about $\pm$ 2 Kelvin units.
}
As usual, one has option of making those intervals more narrow with a reduction of the coverage probability. 

 There can be highly local concerns about environmental issues for which forecasts from one grid cell might target the most relevant spatial unit. For example, rising temperatures affecting a lake or river headwaters can adversely impact the reproductive success of certain species of fish and amphibians. But often, forecasts for a particular grid cell will be of little practical interest. Reported heat waves affect regions that are usually much larger. Policy interventions are needed for the entire region impacted. Moreover, grid cells are not defined by governmental boundaries (e.g., for counties) by which administrative actions are organized. We turn, therefore, to forecasts with days as the temporal units and temperature summaries over all grid cells as the response variable. 

\subsection{Forecasting At The Temporal Level of Days}

We have already computed random forest fitted grid cell values for Q(.95) surface temperatures. A simple and transparent solution to construct a single fitted value for a given day  is to use the mean over grid cells of the random forest fitted value for that day. One can imagine doing something of this sort one day at a time to provide a daily time series. Other summary procedures are available and could be considered in future work.

These ideas can be formalized while retaining the within-subject design. The data format we suggest has a pooled cross-section time series structure with grid cells as cross sections and days as points in time. Such frameworks have been common in the social sciences for decades where they are analyzed as ``panel data'' (Balestra and Nerlove, 1966; Hannan and Young, 1977).  

For any given day,
\[
Y_{i,j} = f(\vec{X}_{(i-14),j}) + \epsilon_{i,j}
\]
\begin{align*}
\text{where:} \quad 
& Y_{i,j} \text{ is the observed Q(.95) surface temperature for day} \hspace{.05in}  i \text{ and grid cell } j, \\
& \vec{X}_{(i-14),j} \text{ is the vector of precursors at time } i-14 \text{ and location } j, \\
& f(\cdot) \text{ is a nonparametric function such as provided by random forests,} \\
& \epsilon_{i,j} \text{ is an error term capturing noise or unobserved effects.}
\end{align*}

The dataset can be in practice initially curated by allowing for many days and many grid cells. The growth could be quite rapid. On the scales we have been using, a month's worth of data over 1000 grid cells would yield 30,000 observations. To arrive at a single fitted temperature value over grid cells of each day, the mean of the grid cell fitted Q(.95) values can be a simple and reasonable summary. 

Accumulating and analyzing such data for forecasting purposes would be a daily task and probably beyond the resources of most practitioners. There would also be challenges to acquire timely data. To take an extreme example, the AIRS data on which we build requires about 12 months to be made available to researchers and will soon be cease to provide any new data at all. The alternative of output from climate simulations does not for us count as primary data and can be compromised by downscaling complications (Gettleman and Rood, 2016; Lafferty and Sriver, 2023). It also has timeliness problems. Areas with relatively dense concentrations of weather stations may have more promise (https://databasin.org/datasets/ 15a31dec689b4c958ee491ff30fcce75/). 

Insofar as the data availability problems are usefully addressed, we offer a computational shortcut. Training would be done once with a convenient form of supervised learning much in the way we proceeded to construct Figure~\ref{fig:fitquality}. One would then step forward (or backward) one day at time with the trained algorithm fixed, but with the new precursor values each day as inputs to obtain the fitted values. The mean of those fitted values is treated as the fitted Q(.95) surface temperature for that day. In other words, once the forecasting algorithm is trained, it can be moved to a different day with that day's precursor values as input to construct fitted values. We are assuming for this application in Siberia that the same trained structure broadly applies and that only the precursor random variables change. This seems to be in the spirit of current climate science analyses of heat waves (Mann, 2018; Mckinnon and Simpson, 2022; Petoukhov et al., 2022; Li et al, 2024); the laws of physics that we are tying to encapsulate do not change from day to day. 

The result would then a daily time series of conditional, mean fitted values serving as the response variable. The only predictor would  be day of the month. We will illustrate this approach shortly with conformal prediction regions computed as well. The pseudocode for analysis of the time series of conditional mean Q(.95) surface temperatures is shown in Algorithm~\ref{alg:conformalQRF}.

\begin{algorithm}[H]
\caption{Forecasts of Daily Q(.95) Surface Temperatures with Conformal Prediction Regions}
\label{alg:conformalQRF}

\begin{algorithmic}[1]
\Require Time series of mean Q(.95) fitted daily surface temperatures 
\ensuremath{\bar{Y}_{i,\bullet}} for days \ensuremath{D_i}, \ensuremath{i=1,2, \dots, T} over grid cells \ensuremath{j=1,2,\dots, N}, miscoverage level \ensuremath{\alpha}, random forests algorithm \ensuremath{\mathcal{F}}, and quantile regression forests algorithm \ensuremath{\mathcal{Q}}.

\Ensure A prediction interval over grid cells for a new, unlabeled observation at \ensuremath{D_{(T+1,\bullet)}}.

\State Apply random forests \ensuremath{\mathcal{F}} to the time series with days as the sole predictor.
\State Compute the fitted values \ensuremath{\widehat{\bar{Y}}_{i,\bullet}} for \ensuremath{i = 1,2, \dots, T}.

\State Compute the random forests residuals:
\[
R_{i,\bullet} = \bar{Y}_{i,\bullet} - \widehat{\bar{Y}}_{i,\bullet}
\]
treated as exchangeable nonconformal scores.

\State Use the random forest predict function to obtain the fitted value \ensuremath{\widehat{\bar{Y}}_{(T+1,\bullet)}} for the new observation \ensuremath{D_{(T+1,\bullet)}}.

\State Fit quantile random forests \ensuremath{\mathcal{Q}} to obtain lower and upper quantile bounds:
\[
\hat{q}_{\alpha/2}\big[D_{(T+1,\bullet)}\big] \quad \text{and} \quad \hat{q}_{1-\alpha/2}\big[D_{(T+1,\bullet)}\big].
\]

\State Return the prediction interval:
\[
\Bigg[\hat{q}_{\alpha/2}\big[D_{(T+1,\bullet)}\big] + \widehat{\bar{Y}}_{(T+1,\bullet)},\; \hat{q}_{1-\alpha/2}\big[D_{(T+1,\bullet)}\big] + \widehat{\bar{Y}}_{(T+1,\bullet)}\Bigg].
\]

\end{algorithmic}
\end{algorithm}

\subsubsection{Forecasts for Late May and Early June 2023}

\begin{figure}[htbp]
\begin{center}
\includegraphics[width=3in]{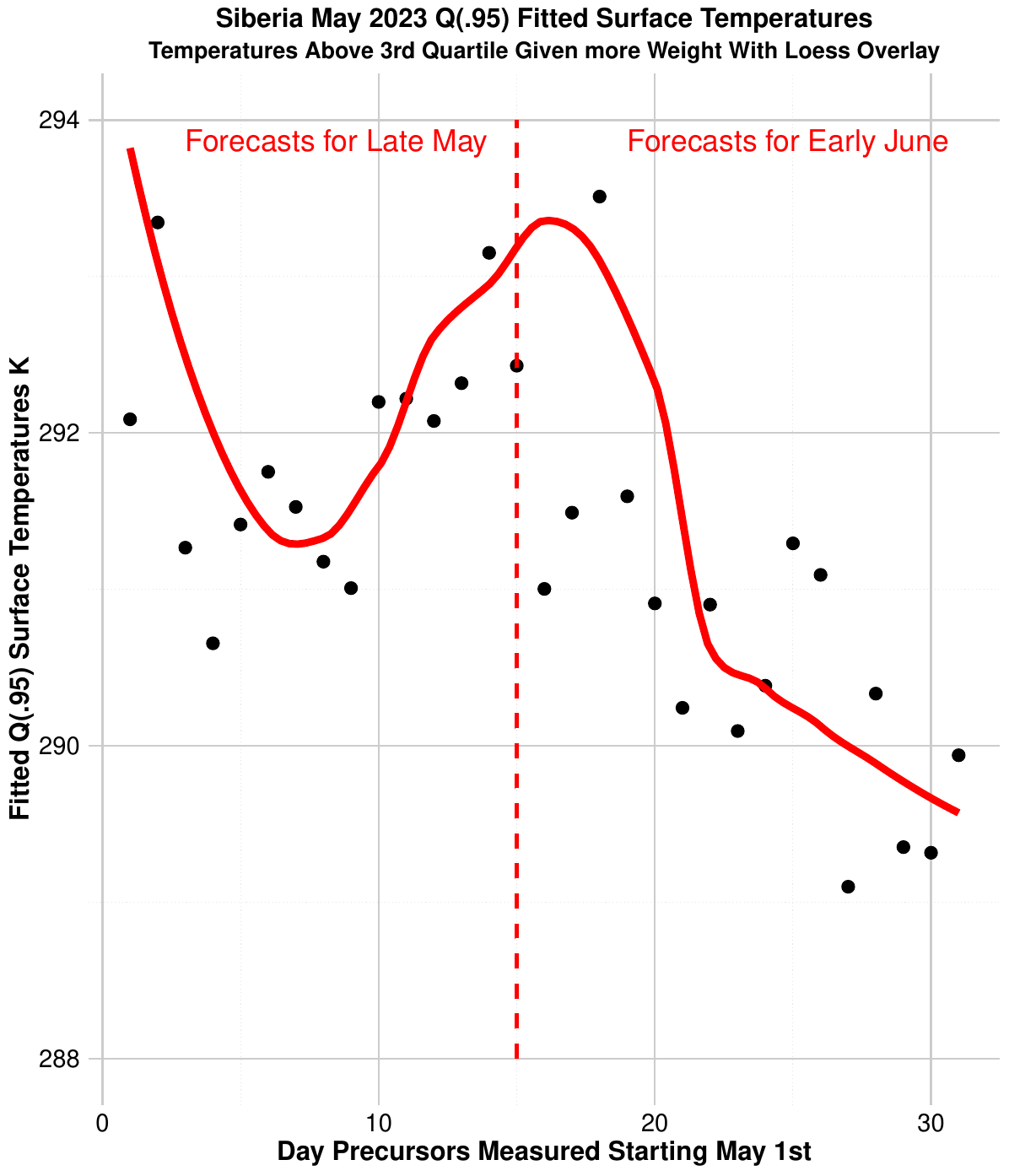}
\caption{Fitted Q(.95) surface temperature values with smoother overlay for May 2023 ``Time stamped'' for precursors used for forecasting average Q(.95) surface temperatures 14 days later. Third quartile temperatures are given more weight in the loess fit.}
\label{fig:overallfit}
\end{center}
\end{figure}

Figure~\ref{fig:overallfit} shows the Q(.95) surface temperature forecasts obtained from the fitted values for the last two weeks in May and the first to weeks in June, an interval that includes the reported 2023 heat wave. A loess smooth is overlaid as a visual aid; it is not a model. Each forecast is ``time stamped'' by the date on which each fitted value's \emph{precursors} were measured. For example, the date May 10th in the figure conveys that the precursors were measured on May 10th, and then used to forecast the average Q(.95) surface temperature for May 24th, 14 days later. The same reasoning applies to each fitted value and the smoother.

Overall, Figure~\ref{fig:overallfit} comports well with earlier results and approximate expectations gleaned from media coverage of the June, 2023 heat wave. The peak immediately to the right of the vertical dashed line that separates forecasts for May from forecasts for June is not surprising because the random forest algorithm was trained with a response variable constructed for very early June. But all of the fitted values for May are \emph{not} extracted directly from the trained results because the surface temperature values being forecasted pre-date the temperatures used in training. The same rationale applies to the Figure~\ref{fig:overallfit} fitted values beginning around the third week in May. These forecasts are projected to June surface temperature that post-date the temperatures used in training.  

\begin{figure}[htbp]
\begin{center}
\includegraphics[width=5.5in]{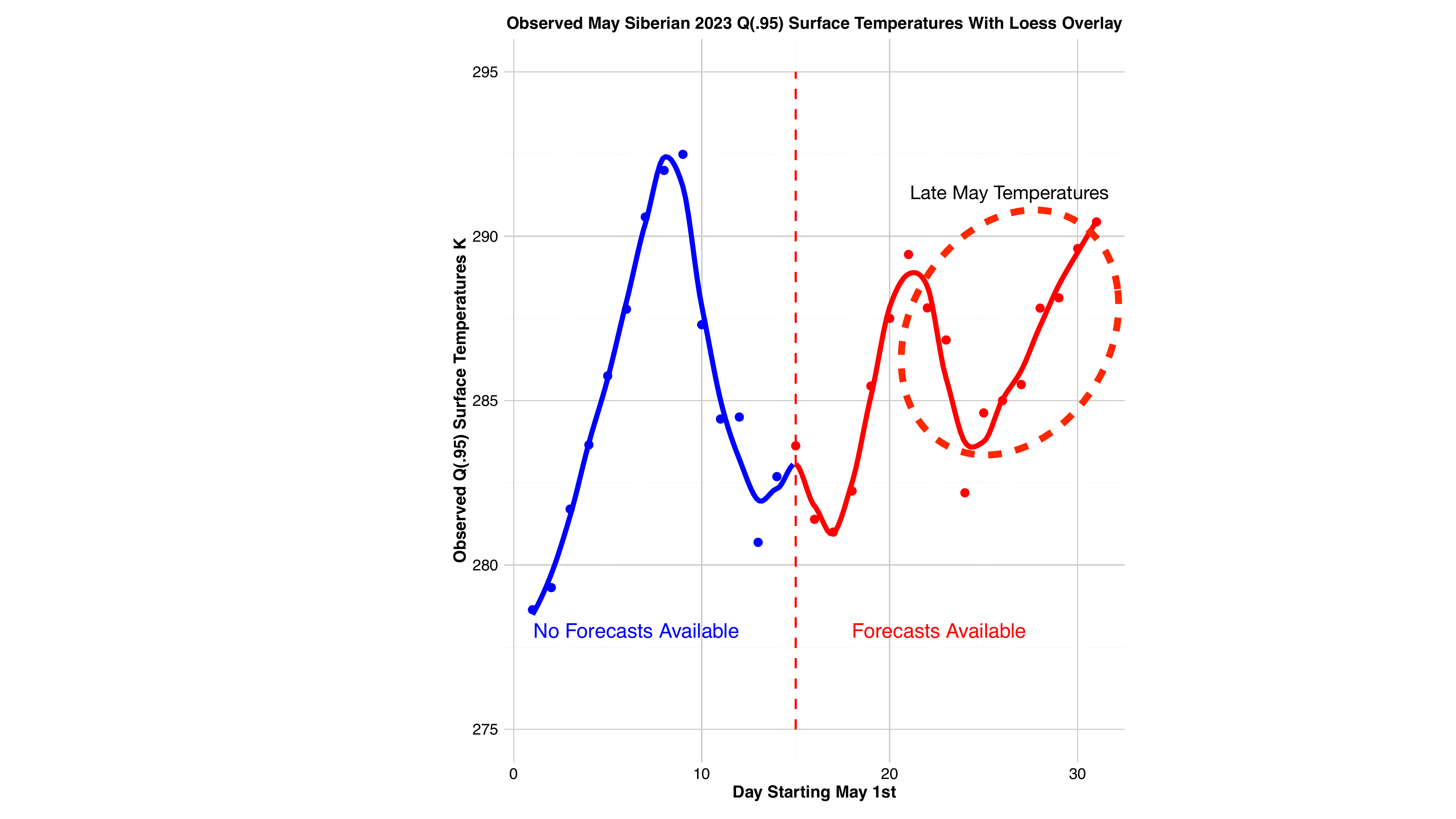}
\caption{Observed May Q(.95) surface temperatures with loess overlay and dashed ellipse corresponding to the late May forecasts in Figure~\ref{fig:overallfit}. The dashed vertical line separates observed temperatures that can be forecasted from the analyses undertaken from those that cannot be forecasted from the analyses undertaken.}
\label{fig:may}
\end{center}
\end{figure}

Figure~\ref{fig:may} shows the observed Q(.95) surface temperatures for May with a loess smoother overly. The portion in blue represents temperatures for which there are no forecasts. Precursor values for April would be required. The portion in red represents the temperature for which forecasts are available in Figure~\ref{fig:overallfit}. The red dashed ellipse highlights the forecasts for May shown in Figure~\ref{fig:overallfit} to the left of the vertical dashed line. The correspondence is substantial especially because it pre-dates the surface temperature used in training. We are also able to forecast quite well the immediately preceding peak. Those forecasts are not shown shown in Figure~\ref{fig:overallfit} because they have little to do with the reported heat wave and would make the figure far more cluttered.

\begin{figure}[htbp]
\begin{center}
\includegraphics[width=5.5in]{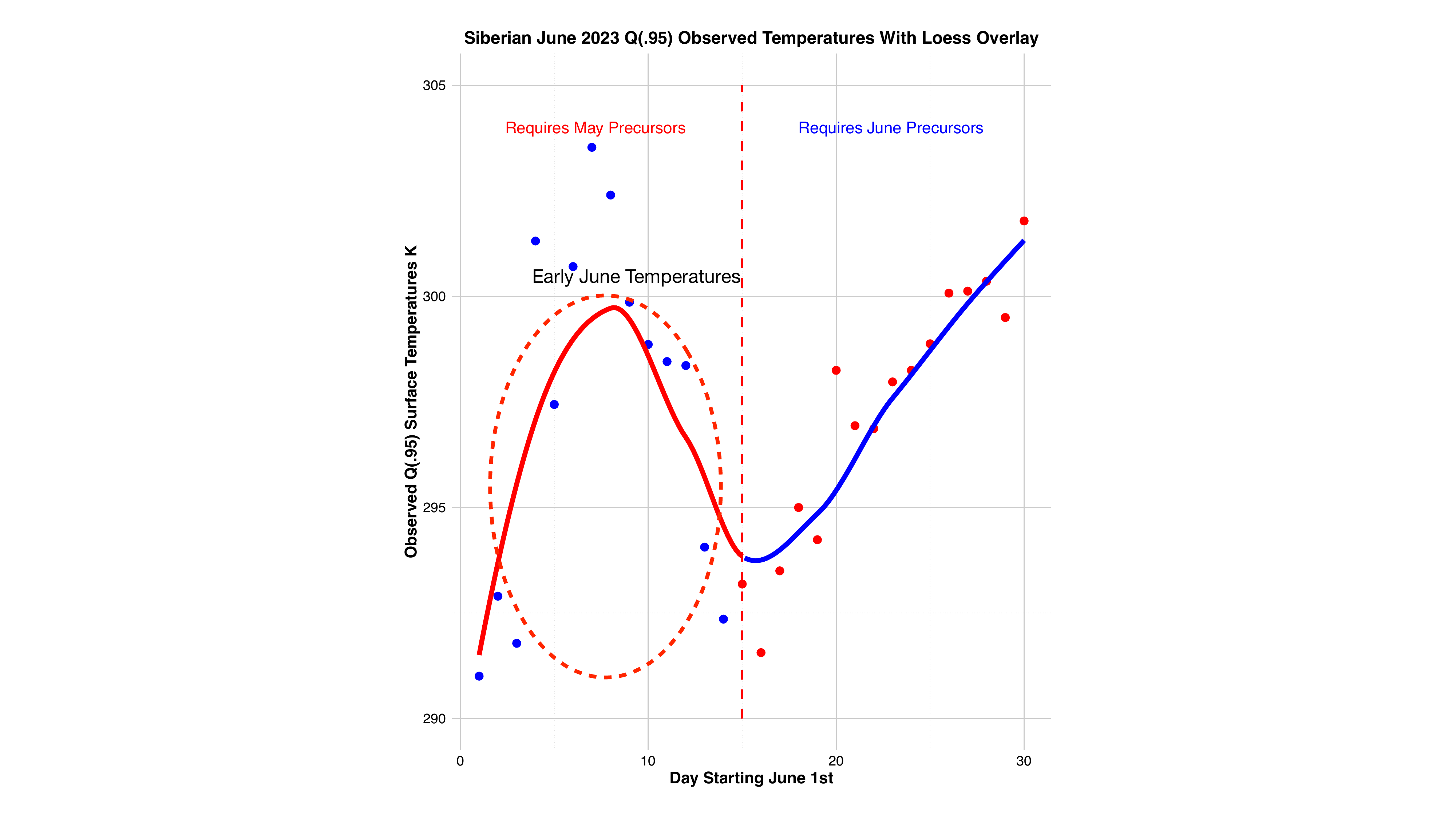}
\caption{Observed June Q(.95) surface temperatures with loess overlay and dashed ellipse corresponding to the early June forecasts in Figure~\ref{fig:overallfit}. The dashed vertical line separates observed temperatures can be forecasted from the analyses undertaken from those that cannot be forecasted from the analyses undertaken.}
\label{fig:june}
\end{center}
\end{figure}

Figure~\ref{fig:june} is constructed in the same manner as Figure~\ref{fig:may}, but for the month of June. During the first two weeks of June there is a trend of rising temperatures followed by a trend in falling temperatures with a turning point at the beginning of the second June week. Looking back at the forecasts on the right side of Figure~\ref{fig:overallfit}, there is a partial correspondence. The forecasted temperatures peak in first three days in June whereas observed temperatures peak near the end of the first week. There is an offset of about 4 days. We suspect that results from the difficulties described earlier determining the date on which the training temperatures were specified.

Figure~\ref{fig:june} also shows to the right of the dashed vertical line the observed Q(.95) surface temperature values for the last two weeks in June. There is a steady and dramatic increase that actually is anticipating a second reported heat wave in early July. We were able to forecast the upward trend in Figure~\ref{fig:june} quite well, but to include that in Figure~\ref{fig:overallfit} would have substantially complicated the graph and addressed temporal structures much later than those associated with the earlier reported June heat wave.

Finally, we provide some illustrative examples of conformal prediction regions for Figure~\ref{fig:overallfit}. An exchangeability assumption for the time series random forest residuals is required, and temporal dependence can be a challenging complication (Chernozhukov et al., 2018). Fortunately, the temporal autocorrelation function for the residuals showed virtually no dependence. The random forest procedure apparently extracted all of the autocorrelation structure in the time series.\footnote
{
It is easy to make the residuals smaller or larger on the average by the amount of smoothness imposed on the fitted values. We had trained the random forests algorithm on the time series data before we tuned and overlayed the loess smoother. But it is likely if we tuned the random forest algorithm to fit the time series less well, temporal structure would have materialized in the residuals. Legitimate holdout data could be very instructive, but for time series data, they would need to come from a time interval not used in the primary analysis. Additional assumptions would follow. 
}

For conformal prediction regions, we proceeded with extensions of tools introduced by Romano and colleagues (2019), adjusted for time series data and making no use of data splitting. With a specified coverage value of .75. Forecasted values were based on the random forest fit. For illustrative new cases the prediction region is within about $\pm$ about .5 degrees K. These conformal prediction regions seem precise enough to be useful at least for policy. The pseudocode mentioned earlier is shown as Algorithm 1.

\section{Discussion}

despite some promising results, missing data remains challenging. Because measurements at different altitudes are often strongly associated, we considered imputing the missing surface temperatures with temperatures from higher altitudes. However, we would  have been imputing values below ground level, which for surface temperatures makes no scientific sense. Had we used values at higher altitudes as if they were surface temperatures, the physical setting would have been rather different and perhaps very misleading. The problem seems inherent in remote sensing data. In this instance, moreover, we would be imputing values for the response variable and building in an artificially improved fit. Future work should consider data collection undertaken in a different manner. Although there are no doubt some spatial sparsity difficulties, progress might be made with weather station data.

Drawing any scientific or policy conclusions for our results unwise. Ideally, the methods we have used might be applied in a number of other sites. Would forecasting accuracy differ? Would similar temporal patterns be found? If similar patterns were found, perhaps the missing data in our analyses are not consequential. A major complication is that temporal patterns for unusually high temperatures may differ legitimately because of variation in topography and climate, not just data quality. The more general point is that forecasting extreme temperatures needs to widely undertaken with timely, complete data that are not synthetic. 

Foundational work on uncertainty quantification must continue. There is a need to supplement conformal prediction regions with methods that do not condition on the training data and trained forecasting procedures. Extensive work is being done on uncertainty assessments that do not depend solely on calibration data, but typically there are other demanding assumptions not easily evaluated with data (Cawse-Nicholson et al., 2020; Braverman et al., 2021). There are also deeper conceptual issues related to discussions surrounding transduction versus induction (Chalvidal et al., 2023); there are tradeoffs between obtaining pristine statistical results in very circumscribed settings compared to untidy statistical results in less controlled settings. One might characterize this as a tension between research reliability and research generalizability. However, if going forward, a trained forecasting procedure will be used ``as is'' and not retrained, the training data and the fitted forecasting procedure are implicitly fixed. No additional uncertainty is introduced by the training data or the trained forecasting procedure, and one need not be concerned about it. When a trained forecasting procedure is provided to hands-on decision-makers, this is often the situation (Berk, 2017).

The problem of endogenous sampling remains. Our methods begin with balanced data, but  we provide the means through weighting to better represent a wide range of real world populations. However, it is well known that weighting can indirectly make loss functions asymmetric because underestimates of certain response variable values can become relatively more (or less) important to the fit. From within a loss minimization framework, this can be undesirable. But as Luce and Raiffa (1957) powerfully argued decades ago, the goal should be risk minimization not loss minimization. Nobel Laureate Clive Granger (1969) made similar points applied to time series data and forecasting. More recent work on the topic includes Waldman (2000), Christoffersen and Diebold (2009), Kuhn and Johnson (2013), Lin et al., 2017),  and Rednick and colleagues (2021). The point is that beginning with highly unbalanced data from an instructive real world population, giving more weight to unusually high measured temperatures would likely better reflect the relevant policy and scientific priorities while producing better balance in the data.\footnote
{
Choosing those weights is ultimately a scientific and policy matter. Expertise beyond statistics and/or computer science is needed. In other settings at least, stakeholders often can quickly arrive at an approximate weighting regime (Berk, 2018). For this paper, such processes are premature.
}

How might this work in practice? An accessible population of interest would need to be specified presumably through concerns about high surface temperatures. Data would then be collected that incorporated a large range of surface temperature values, including the unusually high temperatures that by definition are relatively rare. By weighting more heavily the cases with high temperatures, the risk minimizing algorithm would be encouraged to fit those cases especially well. A byproduct would be a more balanced dataset because the weighting would make the relatively rare high temperature cases less rare. The weights to be used would be a determined by stakeholders. 

As a related matter, implicit in targeting real world populations is a decision about what will serve as one or more comparison settings. These are a key vehicle for determining an appropriate weighting regime. They also go directly to the science as the extraordinary is compared to the ordinary. More should be incorporated than statistical convenience. Most work to date does not clearly identify the comparison setting. For example, despite all of the research on the Pacific Northwest reported heat wave in the summer of 2021, it is commonly unclear to what those processes are being compared. 

We have not addressed how forecasted extreme high temperature are connected to heat waves. Even when efforts have been made to provide precise mathematical/statistical definitions (Perkins and Alexander, 2013; Smith et al., 2013), key heat wave features are justified by empirical norms in climate science and meteorology or computational convenience. Other definitions focus on some empirical characteristic such as a ``heat dome'' (Li et al., 2024) that can provide theoretical connections to climate science, but may be too restrictive. For example, the length of time a heat wave is resident might not be considered part of the heat wave definition. There are also different kinds of heat wave events, depending on such features as humidity, which can have terribly important consequences for public health (Cvijanovic et al., 2023). And each of these approaches should consider whether any current heat wave definition might need revising in a warming climate world. We suspect that eventually a consensus heat wave definition will be provided by theory from climate science, but in the meantime, care must be taken to avoid definitions that unnecessarily restrict the phenomenon's empirical range or misrepresent it fundamentally. In addition, a step back of the sort we have taken might prove helpful. 

\section{Conclusions}

There is no doubt that our results depend substantially on the study site, the time of year, missing data patterns, and perhaps also on special characteristics of the reported heat wave that organized much of the analysis. For example, we only studied forecasted temperature trends; we had no data on humidity, wind direction or wind speed. We also made some judgement calls about the forecasting target, the data structure, and statistical procedures employed. No general claims are being made. Nevertheless, we can place our results in the context of some important current thinking about heat waves.

There is little in our findings consistent with heat waves as discrete events that have sharp beginning and endings. Temperatures seem to build relatively gradually over several days or more and then taper off gradually as well. The gradual warming might be a bit like pre-heating an oven before the baking begins, or it might be seen as part of the heat wave process itself. If more generally, clusters of unusually high temperature build over days or even weeks, the prospects for useful forecasts are substantially improved.

There also seem to be on occasion multimodal peaks a week or more apart. These raise important questions. Should those be treated as one heat wave or more than one? We are returned to the need for advances in climate science and to better inform how to think better about the nature of heat waves.

We found substantial temporal structure in our time series analyses. Our findings depend in part on those differences and underscore the need to always address ``heat wave compared to what?'' Theoretical understanding of heat waves seems to depend fundamentally on explaining, not just describing, how ordinary high temperatures differ from extraordinary high temperatures.

Conformal prediction regions performed largely as anticipated, but take the training data and trained forecasting algorithm as given and, hence, fixed. This  is not entirely satisfactory, but moving to more general assessments of uncertainty probably requires additional and untestable assumptions. In some settings, that may be acceptable if the forecasts are relatively robust to modest assumption violations. In principle, that could be examined. Our use of weak dependence is perhaps an instance.

In summary, the remote sensing observations we employed are not synthetic data produced by climate simulations. The data were curated to form a within-subject research design. Our results are consistent with legitimate temperature forecasts two weeks or more in advance. Valid conformal uncertainty assessments are provided. Q(.95) surface temperatures are forecasted with a suggestive level of skill. Taken at race value, our forecasts imply that several days in a row of projected substantially increasing surface temperatures can be a useful indicator two weeks in advance of an impending set of unusually high surface temperatures that some might call a heat wave.

\section*{References}
\begin{description}
\item
Anastasios N.A., and Bates, S. (2023), ``Conformal Prediction: A Gentle Introduction.'' \textit{Foundations and Trends in Machine Learning} 16(4): 494 -- 591.
\item
 Aumann, H.H., Chahine, M.T.,Gautier, C.,Goldberg, M.D., Kalnay, E.,McMillin, L.M., Revercomb, H., Rosenkranz, P.W., Smith, W.L., Staelin, D.H., and Strow, L.L., ``AIRS/AMSU/HSB on the Aqua Mission: Design, Science Objectives, Data Products, and Processing Systems.'' \textit{Transactions on Geoscience
and Remote Sensing} 41(2): 253 -- 264.
\item
Palestra, P., and Nerlove, M. (1966) ``Pooling Cross Section and Time Series Data in the Estimation of a Dynamic Model: The Demand for Natural Gas.'' \textit{Econometrica} 24(3): 585 --612.
\item
Berk, R. (2017) ``An Impact Assessment of Machine Learning Risk Assessment on Parole board Decisions and Recidivism.'' \textit{Journal of Experimental Criminology} 13: 193 -- 216. 
\item
Berk, R.A., (2018) \textit{Machine Learning Risk Assessments in Criminal justice Settings.} Springer.
 \item
Berk, R.A., Braverman, A., and Kuchibhotla, A.K. (2024) ``Algorithmic Forecasting of Extreme Heat Waves.'' arXiv:2409.18305 [stat.AP]
\item
Braverman, A., Hobbs, J., Teixeira, J., and Gunson, M. (2021). ``Post hoc Uncertainty Quantification for Remote Sensing Observing Systems,'' \textit{SIAM/ASA Journal of. Uncertainty Quantification}. 
\item
Callaghan, T.V., Shaduyko, O., and Kirpotin, S.N. (2021) ``Siberian Environmental Change: Synthesis of Recent Studies and Opportunities for Networking.'' \textit{Ambio} 50(2): 2104 --2107.
  \item
Cawse-Nicholson, K., Braverman, A., Kang, E., Li, M., Johnson, M. and others. (2020). ``Sensitivity and Uncertainty Quantification for the ECOSTRESS Evapotranspiration Algorithm – DisALEXI.'' \textit{International Journal of Applied Earth Observation and Geoinformation} 89: 102088. 
\item
 Chalvidal, M., Serre, T., and VanRullen, R. (2023) ``Learning Functional Transduction.'' \textit{Advances in Neural Information Processing Systems 36} (NeurIPS 2023)
 \item
 Chernozhukov, V., W\"uthrich, K., and Zhu, Y. (2018) ``Exact and Robust Conformal Inference Methods for Predictive Machine Learning with Dependent Data.'' \textit{Proceedings of Machine Learning Research} 75: 1 -- 17
 \item
 Christoffersen, P.F., and Diebold, F.X. (2009) ``Optimal Prediction Under Asymmetric Loss.'' \textit{Econometic Theory} 13(6): 808 -- 817.
 \item
 Cvijanovic, V., Mistry, M.N., Begg, J.D., Gasparrini A., and Rod\'o, X. (2023) ``Importance of Humidity for Characterization and Communication of Dangerous Heatwave Conditions.'' \textit{npj Climate and Atmospheric Science} 6(33).
 \item
 Gettleman, A., and Rood, R.B. (2016)
 \textit{Demystifying Climate Models: A Users Guide to Earth System Models} Springer.
 \item
 Grammerman, A., Vovk, V., and Vapnik, V. (2013) ``Learning by Transduction.'' arXiv:1301.7375v1 [cs.LG]
 \item
 Granger, C.W.J. (1969) ``Prediction with a Generalized Cost of Error Function. \textit{Operational Research Quarterly} 20: 199 -- 207.
 \item
 Hannan, M.T., and Young, A.A. (1977) ``Estimation of Panel Models: Results on Pooling Cross-Sections and Time Series.'' \textit{Sociological Methodology} 8: 52 --83.
 \item
 Hopke, J. E. (2019). Connecting Extreme Heat Events to Climate Change: Media Coverage of Heat Waves and Wildfires. \textit{Environmental Communication} 14(4), 492–508. 
 \item
 Hulme, M., Dassai, S., Lorenzoni, I., and Nelson, D.R., (2008) ``Unstable Climates: Exploring the Statistical and Social Constructions of 'normal' Climate.'' \textit{Geoforum} 40: 197 -- 206.
 \item
 Jacoby, W.G. (2000) ``Loess: A Nonparametric, Graphical Tool For Depicting Relationships Between Variables.'' \textit{Electoral Studies} 19(4): 577 -- 613.
 \item
 Jaque-Dumas, V., Ragine, F., Borgnat, P., Arbry, P., and Bouchet, F. (2022) ``Deep Learning-Based Extreme Heatwave Forecast'' \textit{Frontiers in Climate} 4 - 2022.
 \item
 Jiang, H.,  and Xie, Y.(2024) ``Spatial Conformal Inference through Localized Quantile Regression.'' arXiv: 2412.01098v1.
 \item
Khatana, S.A.M., Szeto, J.J., Eberly, L.Aa, Nathan, A.S., Puvvula, J., and Chen, A. (2024) ``Projections of Extreme Temperature–Related Deaths in the US.'' \textit{JAMA Network Open} 7(9): e2434942. 
\item
Kim, J-H., Kim, S-J., Kim, J-H., Hayashi, M.,and Kim, M-K. (2022) `` East Asian Heatwaves Driven by Arctic-Siberian Warming.'' \textit{Scientific Report} 12, 18025.
 \item
 Klingh\"ofer, D. Braun, M., Br\"uggmann, D.,  and Groneberg, D.A. (2023) ``Heatwaves: Does Global Research Reflect the Growing Threat in the Light of ClimateChange?'' \textit{Globalization and Health} (19)56: 1 -- 17.
 \item
Kuchibhotla, A.K. and Berk, R.A. (2023) ``Nested Conformal Prediction Sets for Classification with Applications to Probation Data.'' \textit{Annals of Applied Statistics} 17(1): 761 -- 785.
\item
Kuhn, M., and Johnson, K. (2013) \textit{Applied Predictive Modeling}. Springer.
\item
Lafferty, D.C., and Sriver, R.L. (2023) ``Downscaling and Bias-Correction Contribute Considerable Uncertainty to Local Climate Projections in CMIP6.'' \textit{npj Climate and Atmospheric Science} 6: 158.
\item
Le Morvan, M., and Varoquaux, G. (2025) ``Imputation for Prediction: Beware of Diminishing Returns.'' arXiv:2407.19804 [cs.AI].
\item
Li, X., Mann, M.E., Wehnerb, M.F., Rahmstorf, S., Petric, S., Christiansena, S. and Carrilloa, J." (2024) ``Role of Atmospheric Resonance and Land–Atmosphere Feedbacks as a Precursor to the June 2021 Pacific Northwest Heat Dome event.'' \textit{PNAS: Earth Atmospheric, and Planetary Sciences} 121(4): 1 -- 7.
\item
Lin, T-Y., Goyal, P., Girshick, R.B., He, K., and Doll\'{a}r, P. (2017) ``Focal Loss for Dense Object Detection.'' arXiv:1708.02002 [cs.CV]
\item
 Mao, H., Maratin, R., and Reich, B.J. (2022) ``Valid Model-Free Spatial Prediction.'' arXiv:2006.15640v2.
 \item
Mann, M.E., Rahmstorf, S., Kornhuber, K., and Steinman, B.A. (2018) ``Projected Changes in Persistent Extreme Summer Weather Events: The Role of Quasi-Resonant Amplification.'' \textit{Science Advances} 4(10) DOI: 10.1126/sciadv.aat3272.
\item
Manski, C.F., and Lerman, S.R. (1977) ``The Estimation of Choice Probabilities from Choise Based Samples.'' \textit{Econometrica} 45(8) 1977 -- 1988.
\item
Manski, C. and McFadden, D. (1981). “Alternative Estimators and Sample Designs for Discrete Choice Analysis,” in C. F. Manski and D. McFadden (eds.) \textit{Structural Analysis of Discrete Data with Econometric Applications.} MIT Press.
 \item
Marx, W., Haunschild, R., and Bornmann, L. (2021) ``Heat Waves: A Hot Topic in Climate Change Research.'' \textit{Theoretical and Applied Climatology} 146: 781 -- 800.
 \item
Maxwell, S.E., Delaney, H.D., and Kelly, K. (2018) \textit{Designing Experiments and Analyzing Data.} Routledge.
\item
McKinnon, K. and Simpson, I.R. (2022) ``How Unexpected Was the 2021 Pacific
Northwest Heatwave?'' \textit{Geophysical Research Letters} 49(18) (e2022GL100380).
 \item
Meinshausen, N. (2006) ``Quantile Regression Forests''. \textit{Journal of Machine Learning Research} 7: 983 -- 999.
\item
Miller, Harvey (2004). ``Tobler's First Law and Spatial Analysis". \textit{Annals of the Association of American Geographers}. 94 (2): 284 -- 289.
\item
Oliveira, R.I., Ornstein, P., Ramos, T., and Romano, J.V. (2024) ``Split Conformal Prediction and Non-Exchangeable Data.'' \textit{Journal of Machine Learning Research} 25:1-38
\item
Papadopoulos, H., Vovk, V., and Gammerman, A. (2011). ``Regression Conformal
Prediction With Nearest Neighbours''. \textit{Journal of Artificial Intelligence Research} 40: 815 -- 840.
\item
Perkins, S.E. (2015) ``A review on the Scientific Understanding of Heatwaves --Their
Measurement, Driving Mechanisms, and Changes at the Global Scale.'' \textit{Atmospheric Research} 165 - 165: 242 -- 267.
\item
Perkins, S.E., and Alexander, L.V., (2013) ``On the Measurement of Heat Waves.'' \textit{Journal of Climate} 26: 4500 -- 4517.
\item
Price, I., Sanchez-Gonzalez, A., Alet, F. et al. (2025) ``Probabilistic Weather Forecasting with Machine Learning.'' \textit{Nature} 63:  84 -- 90. 
\item    
Petoukhov, V., Rahmstorf, S., Petri, S., and Schellnhuber, H.J. (2013) ``Quasiresonant Amplification of Planetary waves and Recent Northern Hemisphere Weather Extremes.'' \textit{Proceedings of the National Academy of Sciences} 110: 5336 -- 5341.
\item
Pitcar, A., Cheval, S., and Frighenciu, M. (2019) ``A Review of Recent Studies on Heat wave Definitions, Mechanisms, Changes, and Impact on Mortality.'' \textit{Forum Geographic} XVII (2): 103 -- 120.
\item
Ridnik, T., Ben-Baruch, E., Zamir, N., Noy, A., Friedman, I., Protter, M., and Zelnick-Manor, L. (2021) ``Asymmetric Loss for Multi-Label Classification.'' \textit{Proceedings of the IEEE/CVF International Conference on Computer Vision} (ICCV), 2021 :82 -- 91.
\item
Romano, Y., Patterson, E., and Cand\`{c}s, E.J. (2019) ``Conformal Quantile Regression.'' In H. Wallach et al., (eds) \textit{Advances in Neural Information Processing Systems}, Volume 32.
\item
Russo, S., Dosio, A., Graverson, R.G., Sillmann, J., Carrao, H., Dunbar, M.B., Singleton, A., Montanga, P., Barbola, P., and Vogt, J. (2014) ``Magnitude of Extreme Heat Waves in Present Climate and Their Projection in a Warming World.'' \textit{Journal of Geophysical Research Atmospheres} 199(22) 12,500 -- 12,512.
\item
Shafter, G., and Vovk, V. (2008) ``A Tutorial on Conformal Prediction." \textit{Journal of Machine Learning Research} 9: 371 -- 421.
\item
Smith, T.T., Zaitchik,  B.F., Gohlke, J.M. (2013) ``Heat Waves in the United States: Definitions, Patterns and Trends.'' \textit{Climatic Change} 118: 811 -- 825.
\item
Thompson, S.K. (2002) \textit{Sampling} Wiley.
\item
Tian, B., Manning, E., Roman, J., Thrastarson, H., Fetzer, and Monarrez, R. (2020)
\textit{AIRS Version 7 Level 3 Product User Guide, version 1.1}. Jet Propulsion Laboratory, California Institute of Technology. Pasadena, CA.
\item
Susskind, J., M. I., Iredell, M.I., Leptoukh, P.J., Kieffer, P. J. L. L.,  Worden, H.M., Gierach, R.R., Hearty, D.F., Chedin, P.M., van de Berg, M.L.B., Wilkins, J.D.L.P, and McMillin, L.S. (2014) ``The Atmospheric Infrared Sounder (AIRS) on the Aqua Satellite: The First Decade of Atmospheric Profiling from Space.'' \textit{Bulletin of the American Meteorological Society} 95(3): 431 -- 442, DOI: 10.1175/BAMS-D-12-00158.1.
\item
Waldman, D.M. (2000) ``Discrete Choice Models with Choice-Based Samples.'' \textit{The American Statistician} 54(4): 303 -- 306.
\item
Witze, A. (2020) ``Why Arctic Fires Are Bad New for Climate Change.'' \textit{Nature} 585: 336 -- 337.
\item
Wood, S.N. (2017) \textit{Generalized Additive Models: An Intrduction with R}, second edition, CRC Press.

\end{description}
\end{document}